# Unveiling Retention Loss Mechanism in FeFETs with Gate-side Interlayer by Decoupling Trapped Charges and Ferroelectric Polarization

Runhao Han, Tao Hu, Jia Yang, Saifei Dai, Yajing Ding, Mingkai Bai, Xianzhou Shao, Junshuai Chai, Hao Xu, Qing Luo, Wenwu Wang, Tianchun Ye, and Xiaolei Wang

***Abstract*—We propose a direct experimental extraction technique for trapped charges and quantitative energy band diagrams in the FeFETs with metal-insulator-ferroelectric-insulator-semiconductor (MIFIS) structure, derived from the physical relationship between $V_{th}$ and gate-side interlayer (G.IL) thickness. By decoupling trapped charges and ferroelectric polarization, we reveal that: (i) The gate-injected charges and channel-injected charges are excessive and maintain consistent ratios to ferroelectric polarization (~170% and ~130%, respectively). (ii) Retention loss originates from the de-trapping of gate-injected charges rather than ferroelectric depolarization. (iii) As the G.IL thickens, the gate-injected charge de-trapping path transforms from gate-side to channel-side. To address the retention loss, careful material design, optimization, and bandgap engineering in the MIFIS structure are crucial. This work advances the understanding of high retention strategies for MIFIS-FeFETs in 3D FE NAND.**

## I. Introduction

Recently, the $HfO_2$-based FeFET with a metal/G.IL/ ferroelectric/channel-interlayer (C.IL)/Si (MIFIS) FeFET structure enhances the MW by incorporating gate-injected charges ($Q_{t_G}$) [1-18]. **Fig. 1(a)** shows that this structure drives the $HfO_2$-based FeFET to meet the requirements of 3D NAND beyond 1K layers, i.e., large MW (>10 V) for 4-bit memory, low operation voltage ($V_{op}$ < 15 V), and acceptable thickness (< 20 nm). Compared to the conventional MFIS structures, the $Q_{t_G}$ in MIFIS structures collaborates with the ferroelectric polarization ($P_s$) and increases the MW [19]. Although previous studies have reported many G.IL designs, including $SiO_2$ [4], $Al_2O_3$ [5], $SiO_2$/SiN/ $SiO_2$ (ONO) [6], $SiO_2$/$HfO_2$/ $SiO_2$ (SHS) [15], SiN [11], and $Al_2O_3$/$HfO_2$/ $Al_2O_3$ (AHA) [3], and achieved an MW up to 19.4 V, there remains a lack of comprehensive guidelines for simultaneously achieving large MWs and robust retention. **Fig. 1(b)** shows that the large MW usually decreases by ~50% after 10 years, especially for the erase state, i.e., $V_{th_ERS}$ [1, 2, 5, 13, 14, 20]. The conventional retention loss (RL) mechanisms considered the depolarization field of FE ($E_{dep}$) and the de-trapping of $Q_{t_G}$ [14]. However, it remains unclear which of these is the dominant factor. To reveal the RL mechanism, the key is to quantitatively obtain the electric field and charge distribution during retention. However, this idea is constrained by the coupling of $P_s$, $Q_{t_G}$, and channel-injected charges ($Q_{t_C}$) in the gate stack.

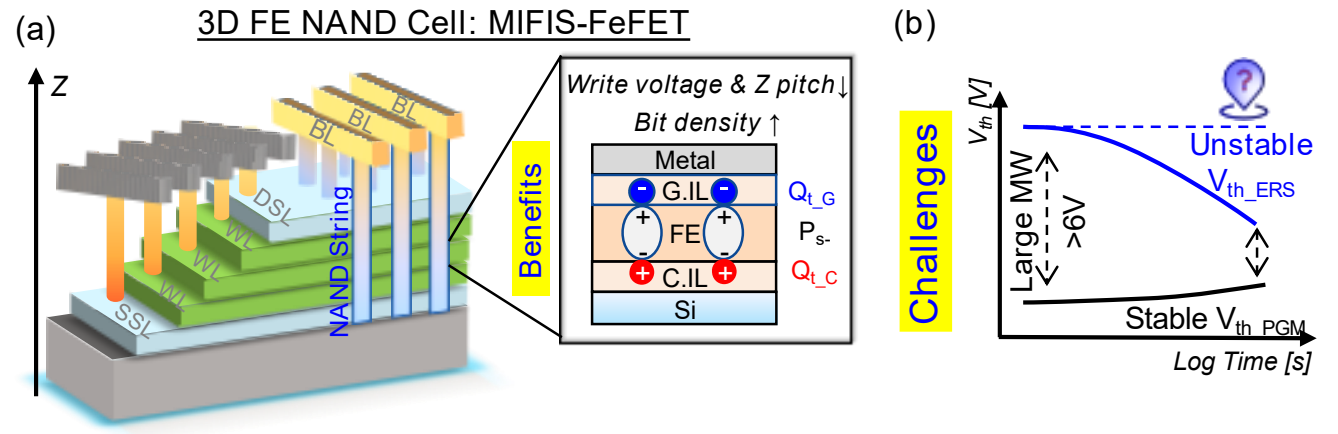

Fig. 1. (a) Schematic of the FE NAND and benefits of the MIFIS structure. (b) The challenges of MIFIS devices: The poor retention mechanism of MIFIS devices is unclear.

In this study, we propose a novel electrical characterization method for decoupling $P_s$, $Q_{t_G}$, and $Q_{t_C}$ and deriving the quantitative energy band diagrams (EBDs) of MIFIS-FeFETs during retention. This is achieved by establishing the $V_{th}$-thickness of G.IL ($d_{G.IL}$) slope as a direct measure of G.IL electric field ($E_{G.IL}$) and combining pulsed *I-V* and *C-V* measurements. We find that the $Q_{t_G}$ is 170% of $P_s$, and the $Q_{t_C}$ is 130% of $P_s$. Both are in an excessive charge injection state. Based on the EBDs during retention, we systematically unveil

This work was supported by the National Natural Science Foundation of China under Grant Nos. 92264104 and 52350195 and Shandong Provincial Natural Science Foundation under Grant No. ZR2022MF313. (Corresponding author: Xianzhou Shao, Junshuai Chai)

Runhao Han, Tao Hu, Saifei Dai, Yajing Ding, Mingkai Bai, Xianzhou Shao, Junshuai Chai, Hao Xu, Qing Luo, Wenwu Wang, Tianchun Ye, and Xiaolei Wang are with Key Laboratory of Fabrication Technologies for Integrated Circuits, Chinese Academy of Sciences, Beijing 100029, China, and also with the Institute of Microelectronics of the Chinese Academy of Sciences, Beijing 100029, China, and also with the School of Integrated Circuit, University of Chinese Academy of Sciences, Beijing 100049, China (e-mail: shaoxianzhou@ime.ac.cn, chaijunshuai@ime.ac.cn).

Jia Yang are with the School of Advanced Interdisciplinary Sciences, University of Chinese Academy of Sciences, Beijing, 101408, China

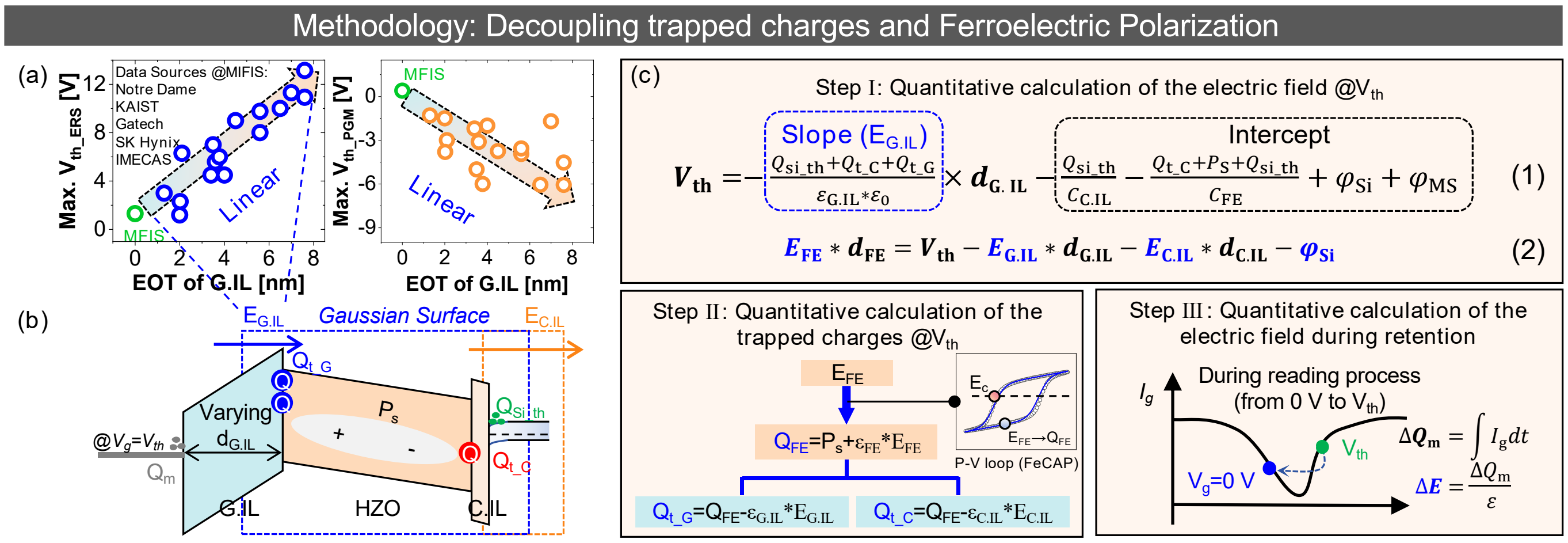

$$V_{th} = -\frac{Q_{si_th}+Q_{t_C}+Q_{t_G}}{\varepsilon_{G.IL}*\varepsilon_0} \times d_{G.IL} - \frac{Q_{si_th}}{C_{C.IL}} - \frac{Q_{t_C}+P_S+Q_{si_th}}{C_{FE}} + \varphi_{Si} + \varphi_{MS} \quad (1)$$

$$E_{FE} * d_{FE} = V_{th} - E_{G.IL} * d_{G.IL} - E_{C.IL} * d_{C.IL} - \varphi_{Si} \quad (2)$$

$$\Delta Q_m = \int I_g dt \qquad \Delta E = \frac{\Delta Q_m}{\varepsilon}$$

Fig. 2. The concept of our method. (a) A linear relationship between EOT of G.IL and Max. $V_{th_ERS/PGM}$ in reported MIFIS devices. (b) The schematic of the energy band diagram for MIFIS-FeFETs at the $V_g = V_{th}$ state. (c) The extraction steps for quantitatively deriving the trapped charges and electric field distribution.

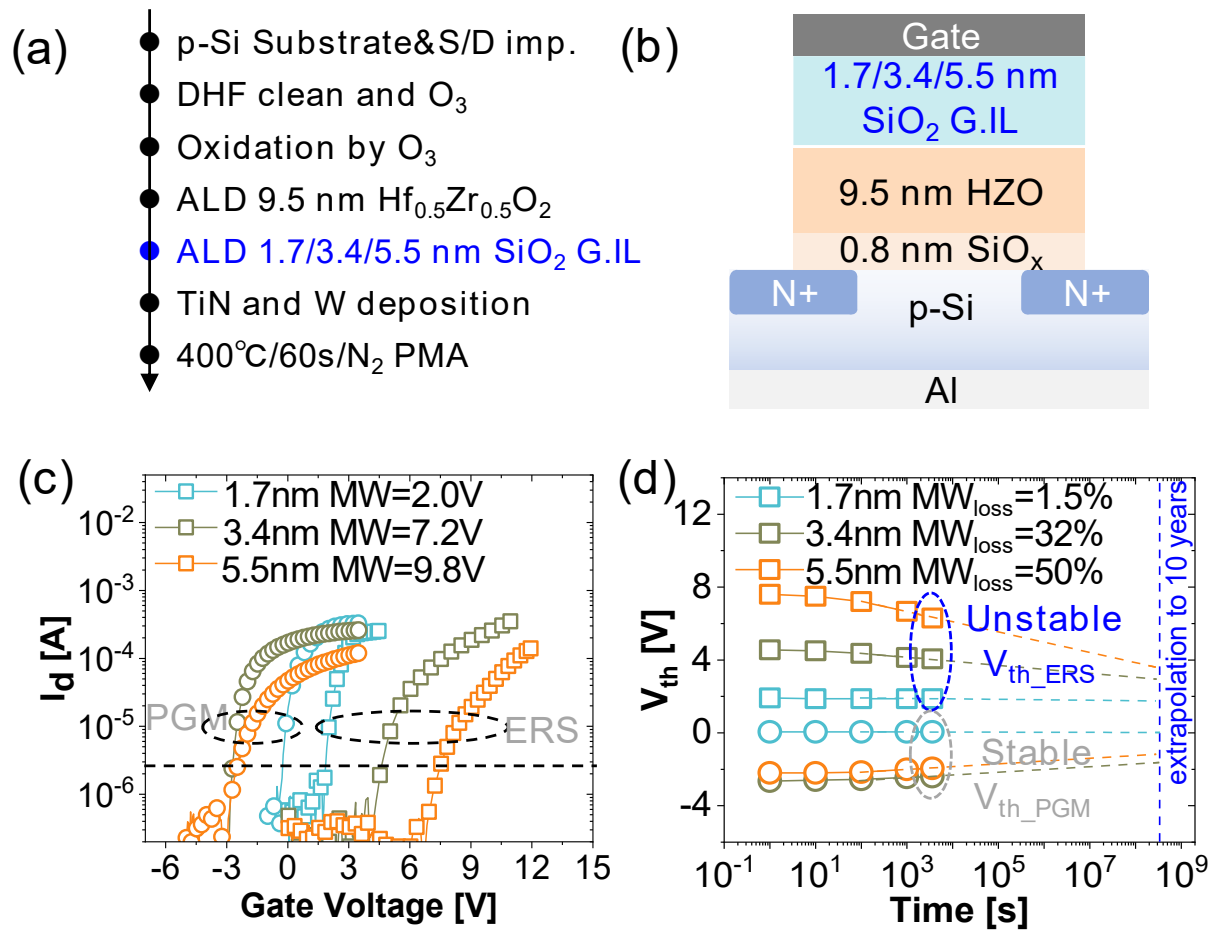

Fig. 3. (a) Fabrication flow. (b) The schematic diagram of three MIFIS-FeFETs with different $d_{G.IL}$. (c) $I_d$–$V_g$ curves of max. MW for all samples. (d) Retention loss of the MIFIS devices with different $d_{G.IL}$. The ERS state is unstable.

that the RL originates from the de-trapping of $Q_{t_G}$ rather than the $E_{dep}$.

## II. Quantitative Extraction Technique

**Figure 2** shows the conceptual framework of this method. The reported results of MIFIS-FeFETs show a universal linear relationship between the maximum $V_{th_ERS/PGM}$ and the EOT of G.IL, as shown in **Fig. 2(a)**. An analytical $V_{th}$ expression (1) for this linear relationship in the MIFIS structure is shown in **Fig. 2(c)**. Guided by Gauss's law, this slope carries a critical physical meaning: the electric field in the G.IL ($E_{G.IL}$) at $V_g$=$V_{th}$ (**Fig. 2(b)**). From this insight, the extraction steps of the electric field and charge distribution in **Fig. 2(c)** are as follows. Step I: obtain the $E_{FE}$ according to equation (2) at the $V_g = V_{th}$ state. The $E_{G.IL}$ can be determined based on the slope, while the $E_{C.IL}$ and surface potential of Si ($\varphi_{Si}$) are determined using the $C_g$-$V_g$ test and Gauss's law. Step II: decouple the $Q_{t_G}$, $P_s$, and $Q_{t_C}$ at $V_g = V_{th}$. The $P_s$ is determined by $E_{FE}$ based on a *P-V* loop obtained from FeCAP. The $Q_{t_G}$ and $Q_{t_C}$ can be calculated using the electric displacement continuity. Step III: derive the electric field and charge distribution at $V_g$ = 0 V, i.e., during retention. The variation in the electric field from 0 V to $V_{th}$ ($\Delta E$) can be derived by monitoring gate charges ($\Delta Q_m$) changes. The variations of $P_s$ and trapped charges from 0 V to $V_{th}$ are negligible [21]. Finally, combining the electric field and charge distribution at $V_g = V_{th}$ and $\Delta E$, the quantitative EBD and charge distribution during retention can be determined.

## III. Results and Discussions

### A. *Quantitative Analysis of Trapped Charges*

We conduct the electric field and charge distribution extraction in MIFIS with 1.7 nm, 3.4 nm, and 5.5 nm $SiO_2$ $d_{G.IL}$. The key process flow and gate stack structure are shown in **Fig. 3(a)** and **(b)**, respectively. **Fig. 3(c)** shows the $I_d$-$V_g$ curves of three devices in the case of the largest MW. The extracted max. MWs with different $d_{G.IL}$ are 2 V (1.7 nm), 7.2 V (3.4 nm), and 9.8 V (5.5 nm), respectively. The retention results of the three devices are shown in **Fig. 3(d)**. RL degradation becomes serious as $d_{G.IL}$ increases, primarily due to the unstable $V_{th_ERS}$. Therefore, the subsequent analysis focuses on the ERS state. In the following discussion, we take the sample with a 3.4 nm G.IL as an example. It should be noted that this method is also applicable to samples with other thicknesses.

Based on the proposed method, $E_{G.IL}$, $E_{C.IL}$, $\varphi_{Si}$, and $P_s$ ($V_{FE}$) are obtained by the direct experimental extraction technique, as shown in **Fig. 4(a)-(c)**. The $V_{G.IL}$ is calculated from the slope (Fig. 4(a)). $V_{C.IL}$ and $\varphi_{Si}$ are calculated from the *C-V* test (**Fig. 4(b)**). By fabricating ferroelectric capacitance with the corresponding thickness and fitting the *P-V* loop, we can extract ferroelectric parameters (**Fig. 4(c)**). Consequently, combining this information, **Fig. 4(d)** shows the quantitative EBD for the 3.4 nm G.IL device at $V_g$=$V_{th}$. **Fig. 4(e)** presents the decoupling of $Q_{t_G}$, $P_s$, and $Q_{t_C}$ for 1.7, 3.4, and 5.5 nm G.IL devices. The results indicate that: (i) The $Q_{t_G}$ and $Q_{t_C}$ maintain consistent ratios to $P_s$ (~170% and ~130%, respectively), highlighting that charge injection exceeds the $P_s$ limit. This differs from the conventional MFIS structure (no G.IL), in which $Q_{t_C}$ is 90% of the $P_s$ [21]. (ii) As $d_{G.IL}$ increases, $P_s$, $Q_{t_G}$, and $Q_{t_C}$ remain unchanged at $V_g$=$V_{th}$. This suggests that the enhanced MW in MIFIS-FeFETs with a thicker G.IL is primarily due to the reduced capacitance of G.IL, rather than the variation of three coupled charges. **Fig. 4(f)** shows the contribution of each charge component to MW based on our experimental decoupling. The contributive magnitude of $Q_{t_G}$ to MW is 4.7-fold that of $P_s$. Therefore, the enhancement of MW is mainly attributed to the $Q_{t_G}$. In contrast, $P_s$ functions to facilitate charge injection and lower operating voltage.

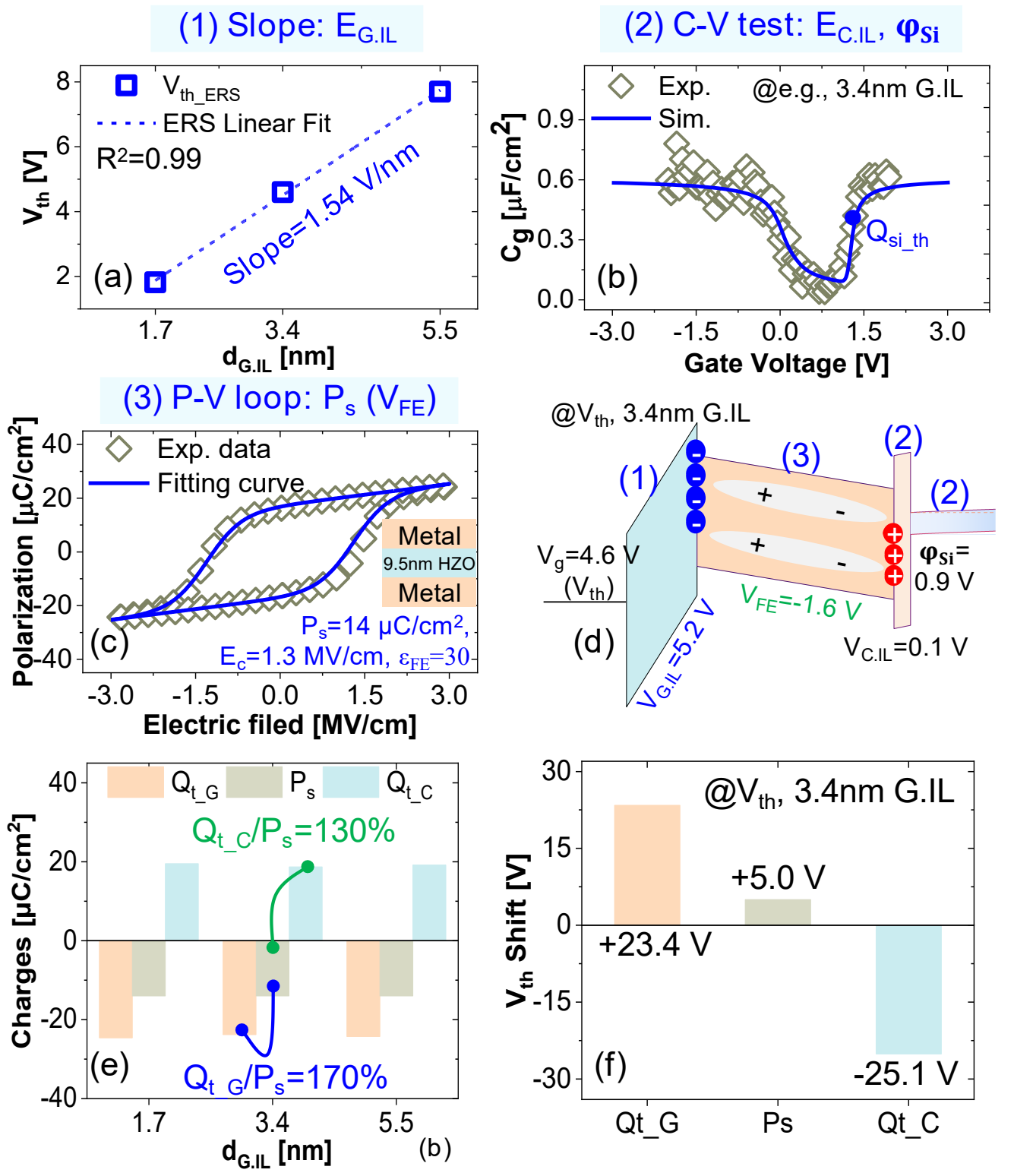

Fig. 4. (a) The linear relationship between the max. $V_{th_ERS}$ and $d_{G.IL}$. (b) Experimental and simulated $C_g$-$V_g$ results of a fresh 3.4 nm G.IL device. (c) *P-V* loop of the corresponding FeCAP and the FE parameters. (d) Quantitative EBD for the 3.4 nm G.IL at $V_g = V_{th}$. (e) The decoupling of $Q_{t_G}$, $P_s$, and $Q_{t_C}$ for 1.7, 3.4, and 5.5 nm G.IL. (f) Contribution of each charge component to the MW.

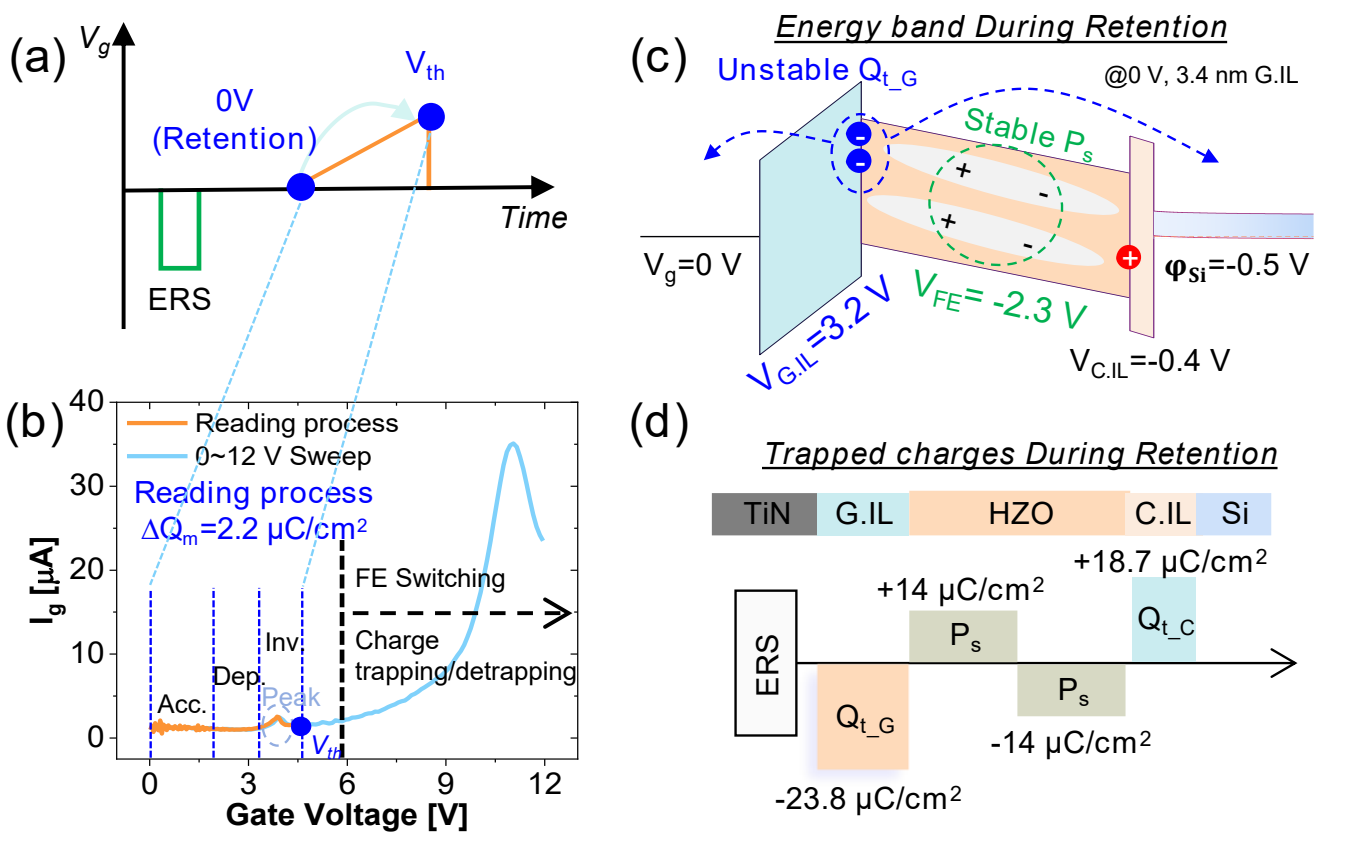

Fig. 5. (a) The gate current ($I_g$) measurement waveform during the reading process. (b) Comparison of $I_g$ results during the reading process (0 ~ 4.6 V) and $I_g$ containing ferroelectric switching and charge trapping/de-trapping (0 ~ 12 V sweep). (c) Quantitative energy band diagram of the 3.4 nm G.IL device during retention. (d) The charge distribution in 3.4 nm G.IL devices during retention.

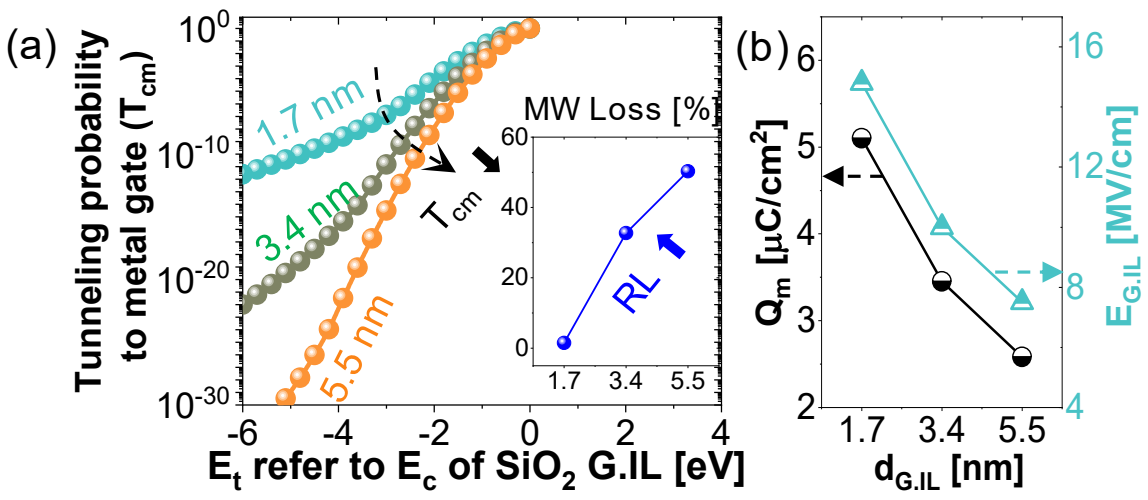

Fig. 6. (a) The tunneling probability for de-trapping of $Q_{t_G}$ to the gate-side ($T_{cm}$) reduces with increasing $d_{G.IL}$. However, retention loss deteriorates with increasing $d_{G.IL}$ (inset). (b) The $Q_m$ decreases with increasing $d_{G.IL}$, thus reducing the $E_{G.IL}$.

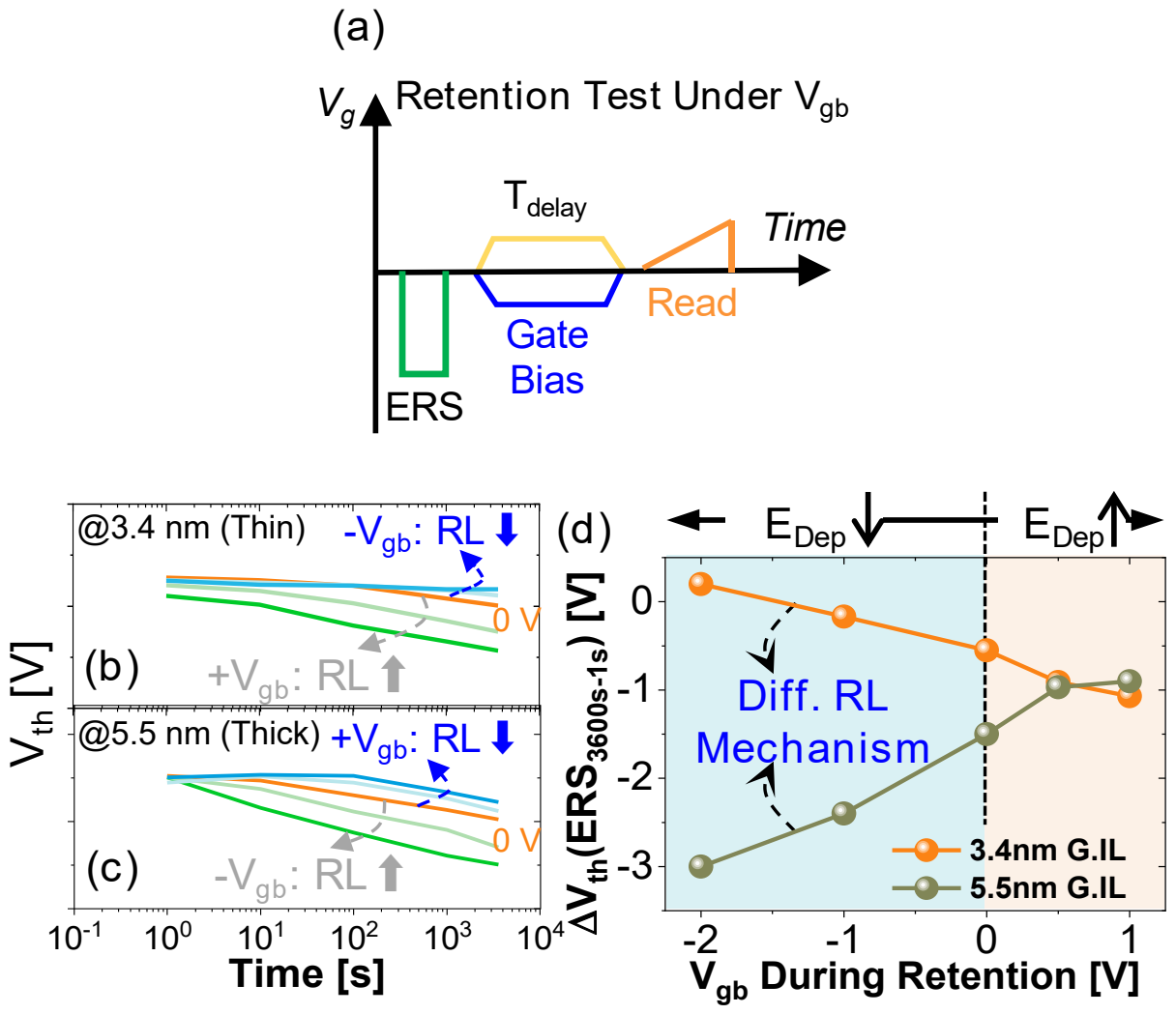

Fig. 7. (a) The test waveform of the retention test under $V_{gb}$. (b) The RL of thin G.IL sample (3.4 nm) under different $V_{gb}$. (c) The RL of thick G.IL sample (5.5 nm) under different $V_{gb}$. (d) The variation in $V_{th_ERS}$ between 1 s and 3600 s under $V_{gb}$.

### *B. Retention Loss Mechanism Analysis*

The variation in $I_g$ during reading operation can be used to calculate the change in charge as $V_g$ transitions from 0 V to $V_{th}$ (i.e., Step III). The electric field and charge distribution at $V_g$ = 0 V can be used to unveil the retention loss mechanism. Therefore, we measure the gate current ($I_g$) during the reading operation. **Fig. 5(a)** shows the waveform during the reading process. **Fig. 5(b)** shows the comparison of $I_g$ results during the reading process (0 ~ 4.6 V) and $I_g$ containing ferroelectric switching and charge trapping/de-trapping (0 ~ 12 V sweep). During the reading process, the $Q_{Si}$ undergoes accumulation (Acc.), depletion (Dep.), and inversion (Inv.). The peak during the reading process is the interface state, which is referred to in [22]. **Fig. 5(b)** shows that no peaks related to FE or charge trapping/de-trapping are observed during the reading process. Therefore, the electric field and charge distribution during retention can be obtained considering that there is only a linear capacitance response during the reading operation [22]. **Fig. 5(c)** and (d) show the quantitative EBD and charge distribution during retention in 3.4 nm G.IL devices. **Fig. 5(c)** shows that the $V_{FE}$ is - 2.3 V, while the $V_{G.IL}$ is + 3.2 V. The negative $V_{FE}$ is beneficial to the stability of polarization. In contrast, the $Q_{t_G}$ is unstable under a positive $V_{G.IL}$ because the energy band bending favors the $Q_{t_G}$ de-trapping to the gate-side or channel-side. This indicates that the RL originates from the de-trapping of $Q_{t_G}$ rather than $E_{dep}$.

Next, we discuss the de-trapping of $Q_{t_G}$. For the $Q_{t_G}$ injected from the metal gate, **Fig. 6(a)** shows that as $d_{G.IL}$ increases, the tunneling probability of $Q_{t_G}$ detrapping to the gate-side ($T_{cm}$) decreases. Here, the tunneling probability was calculated by the WKB approximation. This stems from the lower $E_{G.IL}$ due to the reduced $Q_m$ (**Fig. 6(b)**). However, this result is contrary to the experimental phenomenon of retention loss in **Fig. 6(a)** inset, where the RL degrades with increasing $d_{G.IL}$. Thus, $Q_{t_G}$ detrapping to the gate-side is not the sole mechanism of the RL.

To further investigate the RL, we perform a retention test under gate bias ($V_{gb}$) [14]. **Fig. 7(a)** shows the test waveform. **Fig. 7(b)** shows that, for the thin G.IL sample (3.4 nm), the RL improves under negative bias, while it degrades under positive bias. However, **Fig. 7(c)** shows that, for the thick G.IL sample (5.5 nm), the RL improves under positive bias, and it degrades under negative bias. **Fig. 7(d)** summarizes the variation of $V_{th_ERS}$ shifts between 1 s and 3600 s under $V_{gb}$. The RL results under $V_{gb}$ are opposite in the 3.4 and 5.5 nm samples, indicating different retention loss mechanisms for MIFIS with different $d_{G.IL}$. The mechanism can be explained by energy band changes: Under positive bias, the energy band on the channel-side rises and suppresses the $Q_{t_G}$ detraps to the channel-side. Conversely, the energy band on the metal side rises under negative bias and suppresses the $Q_{t_G}$ de-trapping to the gate-side. Thus, this opposite RL trend indicates that two distinct $Q_{t_G}$ detrapping paths exist.

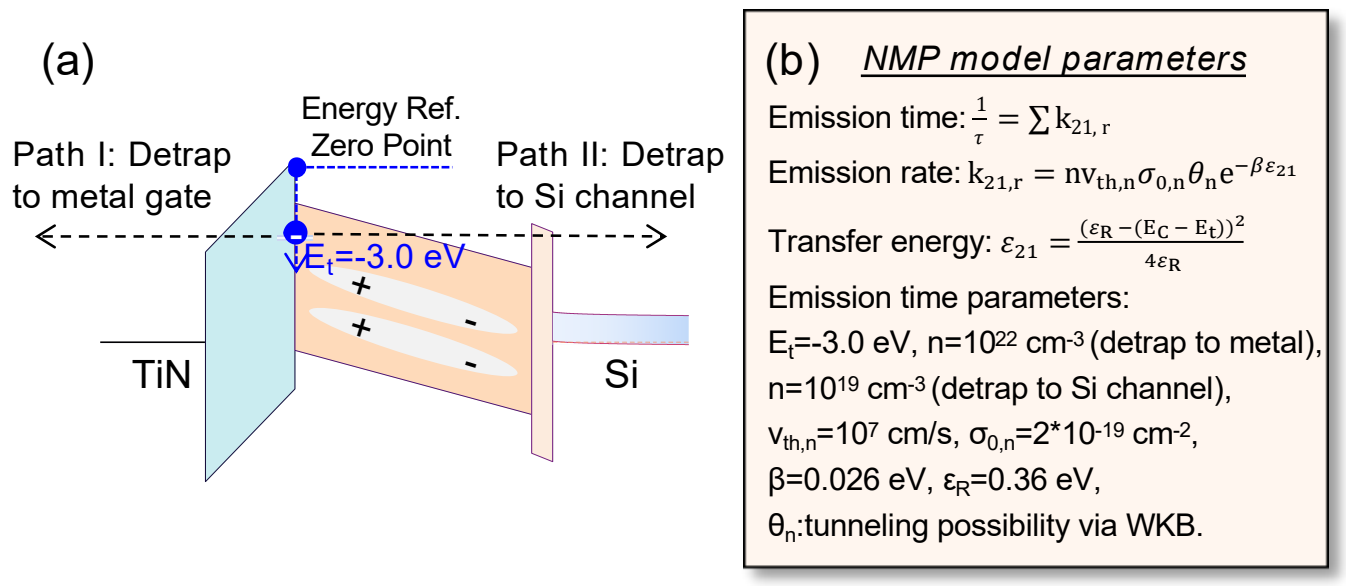

Fig. 8. (a) Two different de-trapping paths of $Q_{t_G}$ for the MIFIS-FeFETs. (b) Nonradiative multiphoton (NMP) model parameters.

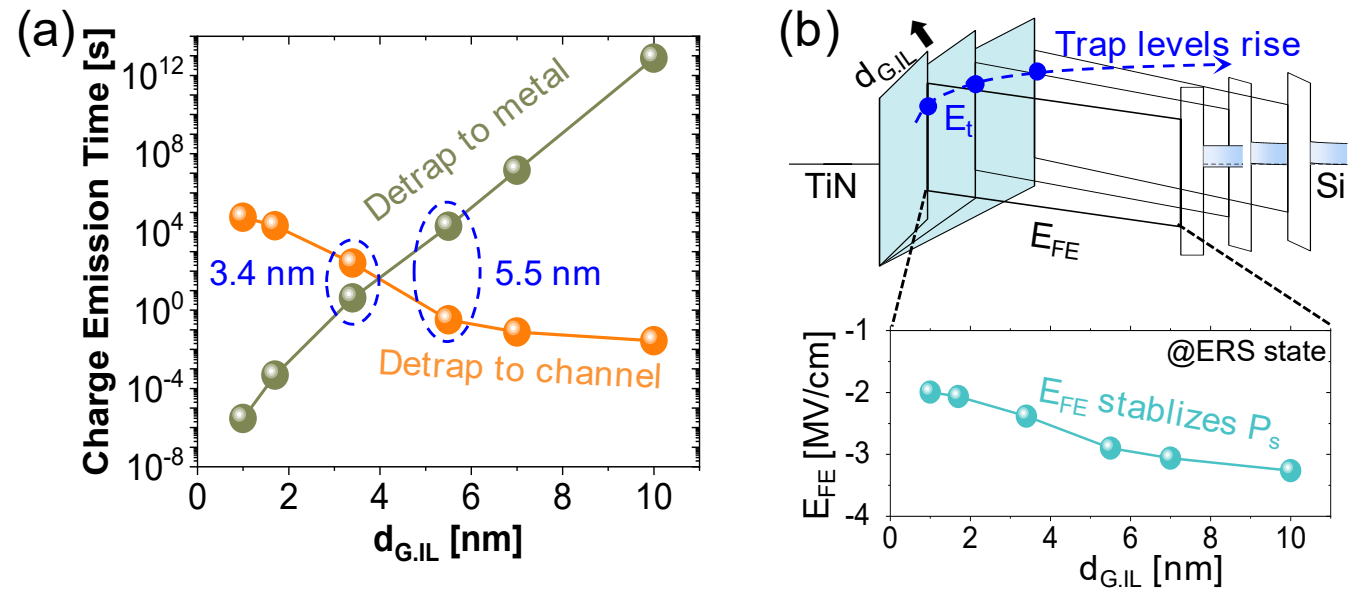

Fig. 9. (a) The charge emission time (Path I and Path II) of different $d_{G.IL}$ samples. (b) The energy band of MIFIS device with different $d_{G.IL}$ during retention. The $E_{FE}$ is more negative with the increasing $d_{G.IL}$.

To further analyze the mechanism between $Q_{t_G}$ de-trapping and $d_{G.IL}$, we establish an RL model based on the EBD during retention. The RL is reflected by the emission time of $Q_{t_G}$. The reference zero point of the $E_t$ is the conduction band of $SiO_2$ G.IL. The $E_t$ is assumed at -3.0 eV [23]. We consider two charge emission paths: to the metal gate-side ($\tau_m$) or the Si channel-side ($\tau_{si}$), as shown in **Fig. 8(a)**. The charge emission process is based on the nonradiative multiphoton (NMP) model, and the parameters are shown in **Fig. 8(b)** [23, 24]. The simulated $d_{G.IL}$ varies from 1 to 10 nm. **Fig. 9(a)** shows the charge emission time results. The $\tau_m$ increases as $d_{G.IL}$ increases from 1 to 10 nm, indicating that the de-trapping of $Q_{t_G}$ toward the gate-side is easier in thin G.IL samples. This arises from a smaller de-trapping barrier at the gate-side in thin G.IL samples under a large positive $E_{G.IL}$. In contrast, the $\tau_{si}$ decreases with increasing $d_{G.IL}$, revealing that the de-trapping of $Q_{t_G}$ to channel-side becomes easier in thicker G.IL samples. **Fig. 9(b)** clarifies the reason: Increasing $d_{G.IL}$ elevates the relative energy level between $E_t$ and the Fermi level of Si, thereby reducing the de-trapping barrier of HZO and C.IL side. Additionally, **Fig. 9(b)** shows that the $E_{FE}$ becomes more negative as $d_{G.IL}$ increases. This further stabilizes the $P_s$.

**Figure 10** delineates the RL mechanism considering two charge de-trapping paths. De-trapping of unstable $Q_{t_G}$ in the ERS state is the primary driver of RL. Two distinct de-trapping paths are identified: toward the gate-side (path I) and toward the Si channel-side (path II). For the thin G.IL samples, path I dominates the RL. For the thick G.IL samples, the path II dominates the RL. This is the reason for the retention loss in the MIFIS-FeFETs with a large MW. Based on this model, we developed the improvement guideline: For a large MW (>10 V), a thick G.IL with low-$\kappa$ is required, considering the capacitance of G.IL. For good retention (> 10 years), a thick G.IL and a large de-trapping barrier for $Q_{t_G}$ are required.

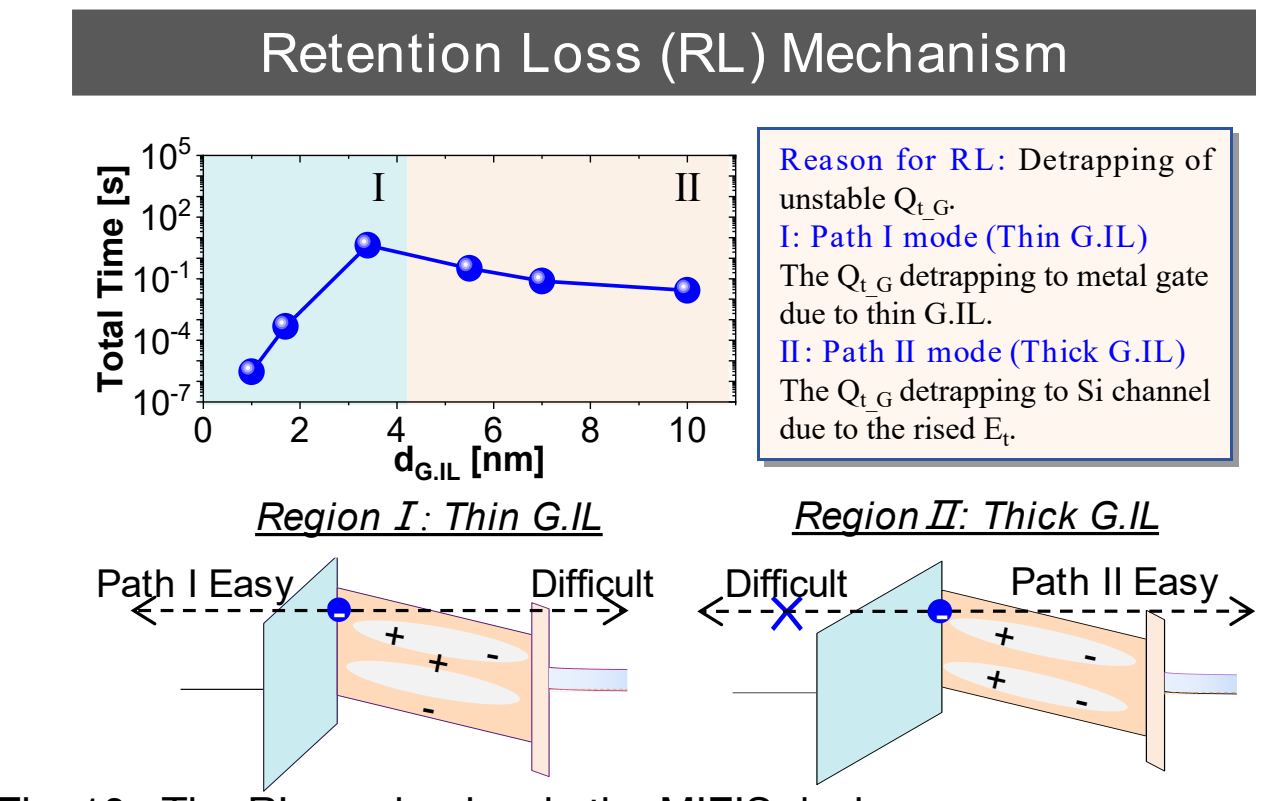

Fig. 10. The RL mechanism in the MIFIS devices.

## IV. Conclusions

The newly developed analysis quantitatively revealed the $Q_{t_G}$ $P_s$ and $Q_{t_C}$ in the MIFIS-FeFET devices. The gate-injected charges and channel-injected charges are excessive and maintain consistent ratios to ferroelectric polarization (~170% and ~130%, respectively). The detrapping of $Q_{t_G}$ is crucial for the retention loss, not the depolarization field. A careful design of the gate stack is required, including precise control of the charge de-trapping barrier, the thickness, and the dielectric constant of G.IL.